\documentclass[pdflatex,sn-mathphys-num]{sn-jnl}

\usepackage{graphicx}%
\usepackage{multirow}%
\usepackage{amsmath,amssymb,amsfonts}%
\usepackage{amsthm}%
\usepackage{mathrsfs}%
\usepackage[title]{appendix}%
\usepackage{xcolor}%
\usepackage{textcomp}%
\usepackage{manyfoot}%
\usepackage{booktabs}%
\usepackage{algorithm}%
\usepackage{aas_macros}
\usepackage{algorithmicx}%
\usepackage{algpseudocode}%
\usepackage{listings}
\usepackage{natbib}%
\usepackage{siunitx}

\newcommand{\micron}{\SI{}{\micro\meter}}

\theoremstyle{thmstyleone}%
\theoremstyle{thmstyletwo}%

\theoremstyle{thmstylethree}%

\newcommand{\STScI}{Space Telescope Science Institute, Baltimore, MD 21218, USA}
\newcommand{\JHU}{Physics and Astronomy Department, Johns Hopkins University, Baltimore, MD 21218, USA}
\newcommand{\UOArizona}{Department of Astronomy/Steward Observatory, University of Arizona, 933 N. Cherry Avenue, Tucson, AZ 85721, USA}
\begin{document}

\title[Article Title]{A New Era for Supernova Science with the \textit{James Webb Space Telescope}}


\author*[1,2]{\fnm{Justin} \sur{Pierel}}\email{jpierel@stsci.edu}
\author[3,1]{\fnm{David} \sur{Coulter}}
\author[4]{\fnm{Christa} \sur{DeCoursey}}
\author[1]{\fnm{James} \sur{DerKacy}}
\author[1]{\fnm{Ori} \sur{Fox}}
\author[1]{\fnm{Melissa} \sur{Shahbandeh}}
\author[1]{\fnm{Matthew} \sur{Siebert}}

\author[1]{\fnm{Michael} \sur{Engesser}}
\author[1]{\fnm{Conor} \sur{Larison}}
\author[1]{\fnm{Thomas} \sur{Moore}}
\author[1,3]{\fnm{Armin} \sur{Rest}}

\author[1]{\fnm{Rodrigo} \sur{Angulo}}
\author[1]{\fnm{Michael} \sur{Dulude}}
\author[4]{\fnm{Eiichi} \sur{Egami}}
\author[1]{\fnm{Massimo} \sur{Griggio}}
\author[3]{\fnm{Mitchell} \sur{Karmen}}
\author[3]{\fnm{Xiaolong} \sur{Li}}
\author[3]{\fnm{Estefania} \sur{Padilla Gonzalez}}
\author[3]{\fnm{Koji} \sur{Shukawa}}
\author[1]{\fnm{Louis-Gregory} \sur{Strolger}}



\affil[1]{\orgname{\STScI}} 

\affil[2]{\orgdiv{NASA Einstein Fellow}}
\affil[3]{\orgname{\JHU}} 

\affil[4]{\orgname{\UOArizona}}



\abstract{Supernovae (SNe) are excellent probes of the physical, chemical, and cosmological context of galaxies across the history of the Universe. The \textit{James Webb Space Telescope} (\textit{JWST}) has a truly unique combination of wavelength, sensitivity, and resolution, providing a new window into SN physics, progenitors, dust production, redshift evolving stellar populations, and cosmology since its launch. Here we summarize the astounding observations \textit{JWST} has made of SNe in its first few years of operations, as well as the corresponding advancements in our understanding of the Universe. This includes detailed views of nearby SNe, the first spectroscopic samples of high-$z$ SNe, and a significant number of gravitationally lensed SNe valuable for both cosmological inference and high signal-to-noise (S/N) spectroscopy. While the upcoming \textit{Nancy Grace Roman Space Telescope} is expected to revolutionize space-based time-domain astronomy, \textit{JWST} and its unique capabilities will continue to play a complementary role by providing high-precision spectroscopic and photometric data for a broad range of SN science cases.  }

\maketitle

\section{Introduction}\label{sec:intro}
Core-collapse supernovae (CC\,SNe), the explosive deaths of massive stars, inject enormous bursts of energy and newly synthesized metals into their environments that drive turbulence, galactic winds, and regulate future star formation \citep{Gentry_SNe_momentum_2020, Gelli_sne_quench_2024,Ibrahim_MZR_evolution_2024}. CC\,SNe are therefore key tracers of star and galaxy evolution,  \citep{Kulkarni_global_metallicity_2005,Madau_SFH_2014}, with measurements of CC\,SN rates as a function of redshift placing constraints on the initial mass function (IMF) and any massive-star-population changes with redshift \citep{Dahlen2004_high_z_SN_rates, Dahlen2012_CCSNe_rates, Strolger2015_CCSNe_rates}. Type Ia SNe (SNe\,Ia), known as ``standardizable'' candles \citep{Hayden2013_SNIa_metallicity, Rigault2013_SNIa_env_dependency, Rubin2020_SNIa_variation, Jones2023_SALT3}, have been used as precise luminosity distance measures to make the initial discovery of dark energy \citep{riess_observational_1998,perlmutter_measurements_1999} and our best local Hubble constant ($H_0$) measurement \citep{riess_comprehensive_2022}.

Since the launch of the \textit{James Webb Space Telescope} (\textit{JWST}), a new frontier of SN science is now accessible. \textit{JWST} has the unsurpassed infrared (IR) sensitivity, aperture, and resolution to push transient discoveries and analyses into an unexplored phase-space of wavelength, depth, and redshift. While \textit{JWST} has a relatively narrow field of view, and therefore does not cover large areas of the sky, it has given us extraordinary new views of SNe compared to previous SN observations afforded by the \textit{Spitzer Space Telescope} \citep[$4$-$25~\mu$m;][]{Fazio2004_Spitzer_IRAC},
and the \textit{Hubble Space Telescope} \citep[\textit{HST}; e.g.,][]{graur_type_2014,rodney_type_2014,strolger_rate_2015}. \textit{JWST} has not only extended the science previously possible with SNe, it has also changed the physical questions we can ask by providing access to new progenitor populations, properties, and explosion conditions across cosmic history.

Here we give an overview of active SNe observed with \textit{JWST}, from detailed analyses of nearby SNe in the rest-frame IR to the first views of extremely high-redshift SNe in the rest-frame ultraviolet (UV)-optical. The paper is organized into discoveries stemming from observations of nearby SNe (Section \ref{sec:local}), \textit{JWST} advancements for cosmology using SNe (Section \ref{sec:cosmo}), and new constraints on the high-$z$ core-collapse SN population (Section \ref{sec:cc}).

\section{Nearby Supernova Breakthroughs}\label{sec:local}
\subsection{Dust Production in Core-Collapse Supernovae}
\label{sub:local-dust}

Dust is one of the key agents shaping the observable universe, influencing everything from the thermal balance of galaxies to the formation of stars and planets. SNe are among the few environments capable of producing dust under the extreme conditions of rapid cooling and high density \citep[e.g.,][]{dwek19}. The composition, quantity, and survival of this dust are directly tied to the chemical enrichment and opacity of the interstellar medium \citep[e.g.,][]{temim15}. Understanding when and how dust forms in SNe, whether early in the dense, cooling gas behind shocks driven by ejecta, circumstellar interaction, or years later as the ejecta cools, provides a window into the origins of cosmic dust seen at high redshift, long before asymptotic giant branch (AGB) stars could contribute significantly \citep[e.g.,][]{michalowski15,boyer25}. Tracing the evolution of dust in SNe, therefore, connects stellar death to galactic evolution and the broader cycle of cosmic material.

\begin{figure*}[t!]
    \centering
    
    \includegraphics[width=5in]{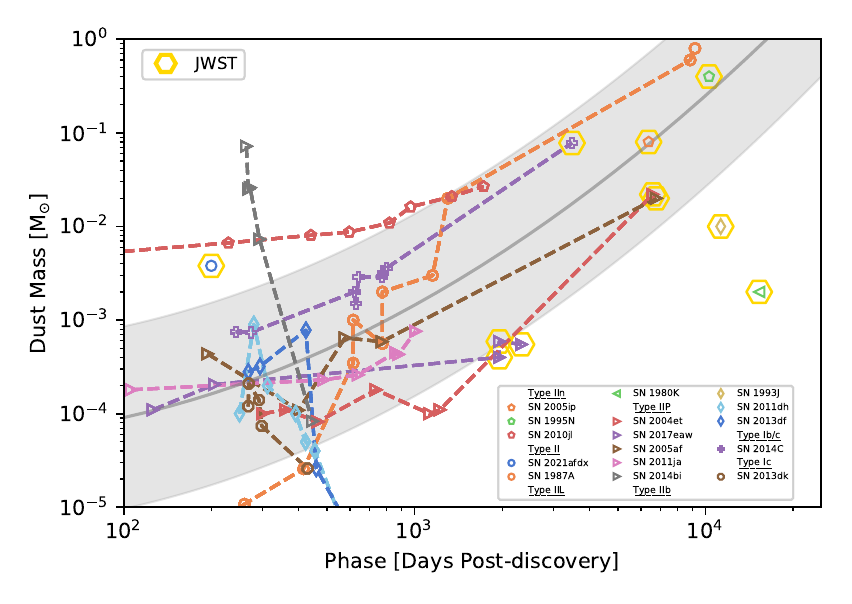}
    \vspace{-10pt}
    \caption{Plot of the {\it inferred} dust mass from mid-IR observations versus time for a number of known dusty SNe. The trend (shaded gray region) empirically suggests an increasing dust mass with time that culminates with significant dust formation by $\sim$30\,yr (e.g., SN 1987A). Data taken from a variety of papers mentioned in the text.}
    \vspace{-10pt}
    \label{fig:dust}
\end{figure*}

Pre-{\it JWST} observations of SNe across the infrared spectrum have revealed that dust formation can occur over a wide range of timescales, from a few hundred days to several years after explosion. Early dust formation is often detected in the dense, cooling ejecta or in regions where the expanding shock interacts with circumstellar material, producing warm dust emission and molecular precursors such as CO and SiO. However, historic dust yields inferred from {\it Spitzer} observations in local core-collapse SNe were generally 2--3 orders of magnitude too small to explain the large amounts of dust observed at high redshifts \citep[e.g.,][and those within]{fox_spitzer_2011,Szalai2019,Szalai21}. That is possibly because many of the {\it Spitzer} observations were obtained during the Warm Mission when only 3.6 and 4.5~\micron\ imaging was possible. These wavelengths are sufficient for tracing the warmest dust (i.e., $>$500 K), but come up short for constraining the dust composition, geometry, and detecting the likely large reservoirs of dust at colder temperatures.

\textit{JWST}/MIRI, with its sensitivity out to 25~\micron, is much more capable of probing new dust formation at early times and uncovering colder dust reservoirs at later times. 
Recent \textit{JWST} observations have provided the first detailed mid-infrared spectra of both early and late phases, enabling direct constraints on dust composition, temperature, and spatial distribution. For example, early observations of SN~2023ixf \citep{derkacy25,medler25,jacobsongalan25} and SN~2024ggi \citep{baron25,mera25} have revealed strong signatures of molecules such as CO weeks to months after explosion. These molecules act as key cooling agents, allowing the ejecta to reach temperatures where dust can begin to condense. Their rapid formation and persistence indicate efficient chemistry and mark the earliest stages of dust production. Ongoing \textit{JWST} and ground-based observations are now capturing the transition from molecular gas to solid grains, offering the clearest view yet of how newly exploded stars begin seeding their environments with dust.

\textit{JWST}/MIRI imaging has also detected some of the largest-ever SN dust masses more than a decade post-explosion \citep[Figure \ref{fig:dust}; e.g.,][]{shahbandeh22,hosseinzadeh23,gomez24,zsiros24,pearson25,sarangi25,szalai25,clayton25,tinyanont25,shahbandeh25} aside from SN 1987A \citep{jones23,arendt23,bouchet24,matsuura24,larsson25}. These nearby SNe span a range of subclasses. Assuming the dust origin and heating mechanism can be constrained \citep{fox10}, these objects provide rare opportunities to directly measure the total dust yield from core-collapse SNe once the ejecta have cooled and expanded sufficiently for mid-infrared emission to emerge unobscured. These observations demonstrate that significant dust production and evolution continue for many years after the initial explosion, underscoring the role of SNe as persistent and efficient contributors to the cosmic dust budget. \textit{JWST}/MIRI medium resolution spectroscopy (MRS) of these systems can further characterize the dust composition (silicate or carbonaceous), geometry, and optical depth. In some cases, the results have shown that the dust is forming in the post-shock region instead of the ejecta, which may be a key to survival \citep[e.g.,][]{shahbandeh25}. All together, these studies suggest that SNe can produce substantial dust masses that survive, potentially accounting for the large dust reservoirs observed in the early universe.

\subsection{Progenitors of Type Ia Supernovae}
\label{sub:local-ia}

\begin{figure*}[!h!]
    \centering
    \includegraphics[width=4in]{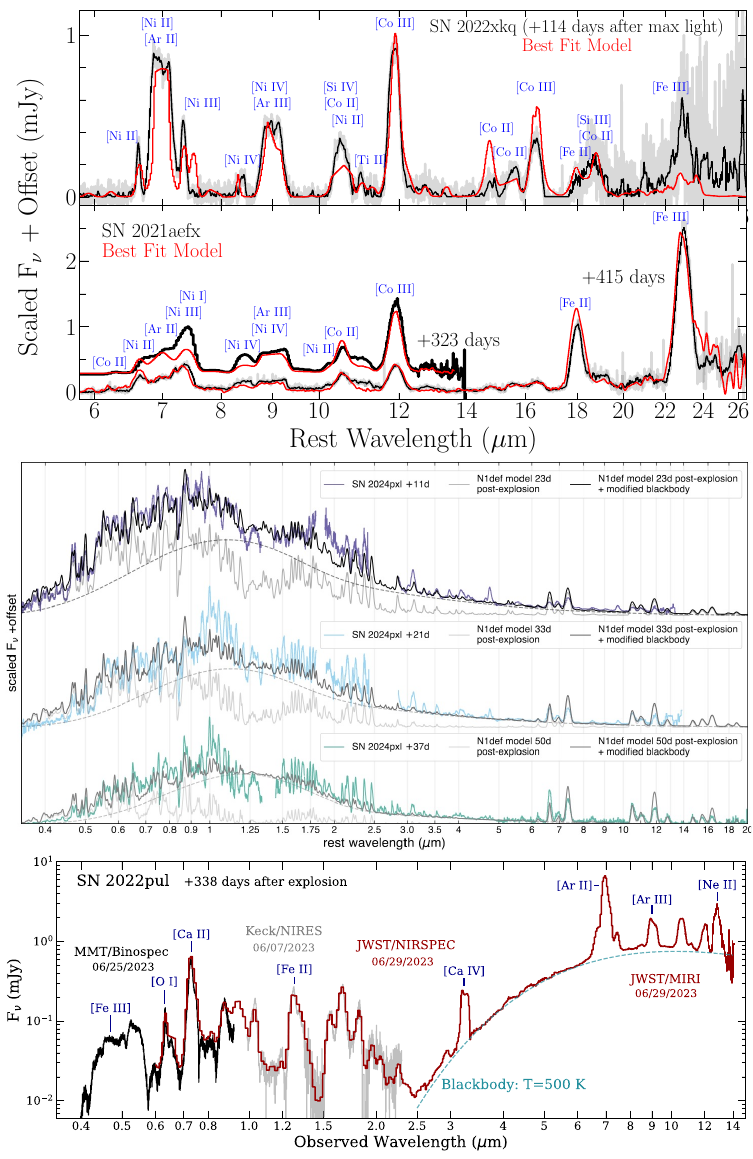}
    \caption{\textit{Top}: \textit{JWST} MIR spectra of SNe~2021aefx and 2022xkq
    compared to 3-dimensional NLTE models of delayed-detonation SNe~Ia models
    from \cite{Hoeflich2017}.
    Strong features are identified. Adapted from \cite{DerKacy2023,DerKacy2024,Ashall2024}.
    \textit{Middle}: Panchromatic observations of SN~2024pxl compared to the 
    1D NLTE synthetic spectra of the N1def model \cite{Fink2014} with and without 
    an extra 4000~K blackbody continuum added. Adapted from \cite{Kwok2025_Iax}.
    \textit{Bottom}:
    Combined optical $+$ near-infrared (NIR) $+$ mid-infrared (MIR) spectrum of the ``Super-Chandra" Type Ia SN (SN\,Ia) 2022pul in the nebular phase at $+338$ days after explosion \citep{Siebert24a}. With optical data from MMT/Binospec at a similar phase, along with slightly earlier NIR data from Keck/NIRES, the \textit{JWST}/NIRSpec and \textit{JWST}/MIRI data continuously cover 0.35--14~$\mu$m, and are the first MIR spectra of a ``Super-Chandra" SN Ia.}
    \label{fig:jwst_localsneia}
\end{figure*}

Prior to the launch of \textit{JWST}, observational constraints on the 
behavior of SNe~Ia beyond wavelengths obtainable from ground-based 
near-infrared observations ($\lambda \gtrsim 2.5$~$\mu$m) were 
limited to a handful of photometry points 
\citep[e.g.,][]{McClelland2013,Fox2013,Johansson2017,Szalai2019} 
and spectra \citep{Gerardy2007,Telesco2015}. Nebular-phase 
observations in the near- and mid-infrared hold immense promise 
for answering several outstanding questions regarding SNe~Ia, 
including: the nature of the circumstellar material in the progenitor 
system \citep{Patat2007,Wang2008,Maeda2015}, whether SNe~Ia can 
form molecules \citep{Hoeflich1995} or dust \citep{Nozawa2011}, how 
flux redistributes in the ejecta with time \citep{Fransson2015}, 
and determining the progenitor white dwarf and explosion properties 
through the presence of isotopes of stable iron-group elements 
\cite{Hoeflich1996,Timmes2003,Seitenzahl2017,Blondin2022,Ruiter_Seitenzahl25_rev}. 
Early observations with \textit{JWST} have primarily focused on obtaining 
nebular-phase spectra for the diversity of SNe~Ia subtypes 
\citep[e.g.,][]{Taubenburger2017}. The normal-bright SN~2021aefx showed 
broad stable [Ni] features \cite{Kwok2024}, indicative of 
high-density burning in a massive ($M > 1.2 M_{\odot}$) white 
dwarf \citep{DerKacy2023,Ashall2024}. Observations of SNe~2022aaiq 
and 2024gy reveal narrow cores of [Ni~II] in addition to the broader
features, suggestive of enhancements of stable Ni in the inner 
$v_{\rm{ejecta}}\sim1000$~km~s$^{-1}$ of the ejecta \cite{Kwok2025_mrs}.
The underluminous SN~2022xkq also 
showed narrow, centrally located [Ni] lines \citep{DerKacy2024}. 
Non-local thermodynamic equilibrium (NLTE) spectral modeling of both SNe~2021aefx and 2022xkq shows good 
agreement with delayed-detonation models \citep{DerKacy2023,DerKacy2024,Ashall2024}.
Optical through mid-infrared 
observations of SNe~2024pxl and 2024vjm, both members of the Iax 
subclass, taken $\sim 10-45$~days after maximum light simultaneously 
show P Cygni and nebular lines, and are well matched by spectral 
models of weak deflagrations \citep{Kwok2025_Iax}. Finally, panchromatic 
observations of the 03fg-like SN~2022pul reveal a strong thermal 
dust continuum beyond 2.5~$\mu$m, with the presence of [O~I] and 
[Ne~II] emission and large scale asymmetries in some line profiles 
supporting origins from a violent merger within C/O-rich circumstellar medium (CSM) 
\citep{Kwok2024,Siebert2024}. Data and models for many of these
objects can be found in Figure \ref{fig:jwst_localsneia}.
Future work will extend the methodologies of these analyses 
to cover the complete time-evolution of SNe~Ia and the full 
phase-space of SNe~Ia subtypes.

\

\section{Supernovae used for Cosmology}\label{sec:cosmo}
\subsection{A Revolution for Strongly Lensed Supernova Cosmology}
\label{sub:cosmo-lensed}
\begin{figure}[h!]
    \centering
    \includegraphics[width=\linewidth]{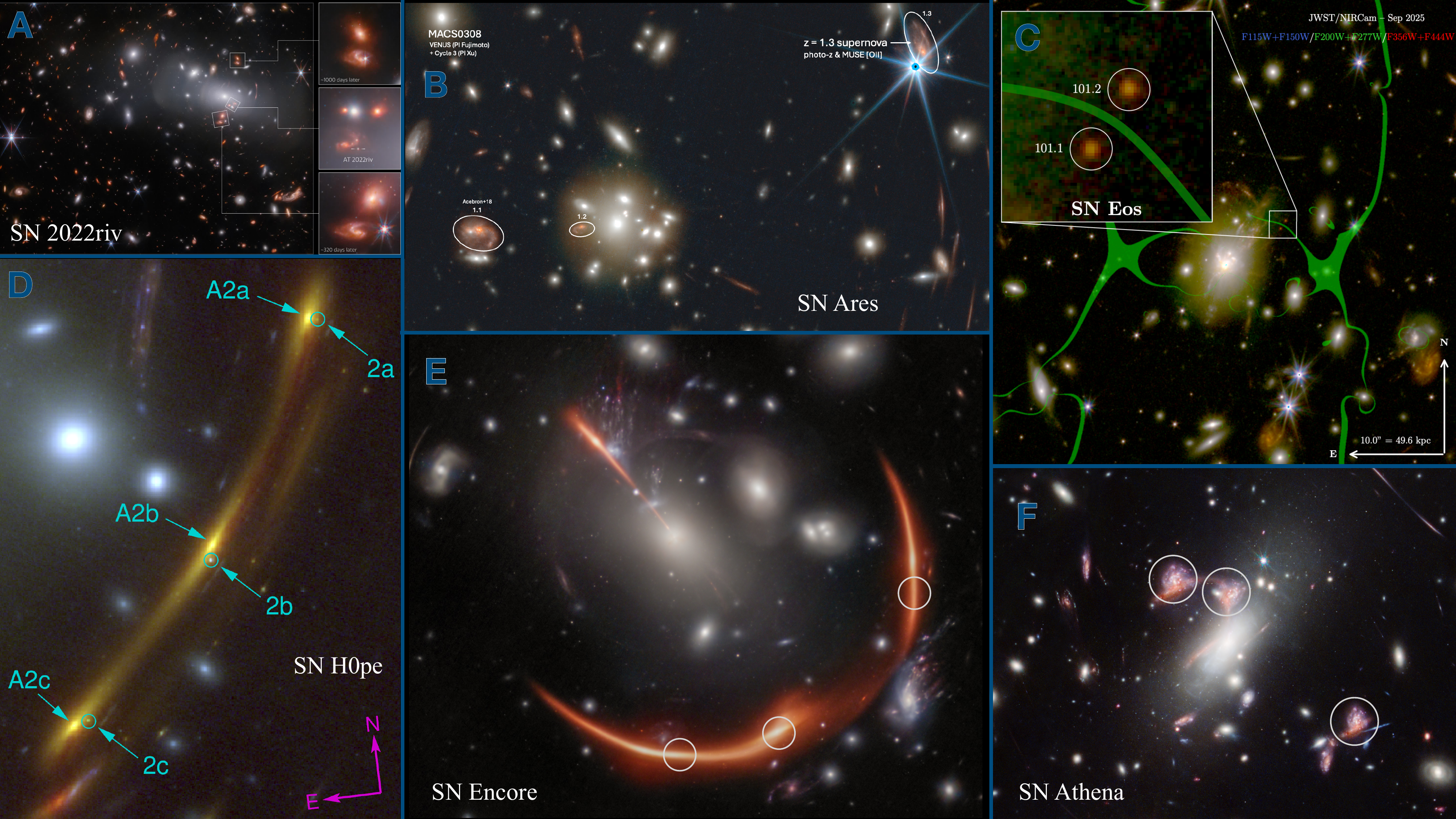}
    \caption[Caption for LOF]{The \textit{JWST} lensed SN sample thus far. (A) SN\,2022riv is a triply-imaged SN\,Ia at $z=1.52$, but only the last image to arrive was detected \citep{kelly_strongly_2022}. (B) SN\,Ares is a triply-imaged SN\,II at $z=1.28$, with the first image to arrive detected and up to two future reappearances in $\sim50+$ years\footnotemark. (C) SN\,Eos is a quintuply-imaged SN\,II at $z=5.13$ with the last two images to arrive (with small time delays) observed by \textit{JWST} (see Section \ref{sub:highz-eos}), image adapted from \citet{coulter_spectroscopically_2026}. (D) SN\,H0pe is a triply-imaged SN\,Ia at $z=1.78$, and was used for the first $H_0$ measurement leveraging a lensed SN\,Ia \citep{pascale_sn_2025}. Figure adapted from \citet{frye_jwst_2024}. (E) SN\, Encore is a multiply-imaged SN\,Ia at $z=1.95$ \citep{pierel_lensed_2024,dhawan_spectroscopic_2024}, with the first three images detected and one or two future images to arrive \citep[depending on the lens model,][]{suyu_cosmology_2025}, which produced another measurement of $H_0$ \citep{pierel_cosmology_2025,suyu_cosmology_2025}. Figure adapted from \citet{pierel_lensed_2024}. (F) SN\,Athena is an unclassified triply-imaged SN at $z=0.87$, with the second image to arrive detected and another image expected in $\sim1$ year\footnotemark[2].}
    \label{fig:lensed_sne}
\end{figure}

When a SN is strongly lensed by a foreground galaxy cluster, multiple images of the SN appear with offsets in time \citep[e.g.,][]{kelly_multiple_2015}. These ``time delays'' \citep{refsdal_possibility_1964} depend both on the lens total mass distribution and on the time-delay distance \citep{suyu_dissecting_2010}, which is inversely proportional to the Hubble constant ($H_0$). 
Thus, strongly lensed SNe are a purely geometrical probe of $H_0$ which, crucially, is fully independent from local distance ladder and early-Universe measurements. Before \textit{JWST}, and over the course of almost a decade, SN Refsdal \citep{kelly_multiple_2015} was the only multiply-imaged SN to yield an $H_0$ measurement \citep[with $\Omega_m=0.3$: 64.8$_{-4.3}^{+4.4}$ or 66.6$_{-3.3}^{+4.1 }$ km s$^{-1}$ Mpc$^{-1}$, depending on lens model weights;][]{kelly_constraints_2023,kelly_magnificent_2023,grillo_cosmography_2024}, despite several other discoveries \citep{goobar_iptf16geu:_2017,rodney_gravitationally_2021, chen_shock_2022,goobar_uncovering_2023,pierel_lenswatch_2023}. \textit{JWST} has revolutionized the field with the discovery of six more cluster-scale lensed SNe in the last $\sim$two years (Figure~\ref{fig:lensed_sne}).

Of these discoveries, two have already provided new $H_0$ measurements \citep[SN~H0pe ($H_0=75.7^{+8.1}_{-5.5} \ \rm{km} \ \rm{s}^{-1}\rm{Mpc}^{-1}$) and SN~Encore ($H_0=66.9^{+11.2}_{-8.1} \ \rm{km} \ \rm{s}^{-1}\rm{Mpc}^{-1}$);][]{frye_jwst_2024,pierel_jwst_2024,pierel_lensed_2024,ertl_cosmology_2025,pascale_sn_2025,pierel_cosmology_2025,suyu_cosmology_2025}, tripling the cosmology sample in a very short time-frame. Each of these $H_0$ measurements was made using photometry that did not include reference imaging (i.e., no difference imaging was possible), and therefore these precisions are expected to increase in the near future \citep{frye_sn_2024,suyu_sn_2025}. Three more of these new lensed SNe have future reappearances (SNe Athena, Encore, and Ares) in $\sim1$, $6$, and $70$ years respectively, providing opportunities to predict the SN arrival times and locations for rare lens modeling systematics checks. Further, SN Encore shares a host galaxy with the previously discovered SN Requiem \citep{rodney_gravitationally_2021}, which is expected to reappear in $\sim2026$ \citep{suyu_cosmology_2025} and will be observed with \textit{JWST} \citep{pierel_requiems_2025}.

\subsection{Testing for Type Ia Supernova Evolution}
\label{sub:cosmo-ia}
\begin{figure}[h!]
    \centering
    \includegraphics[width=\linewidth]{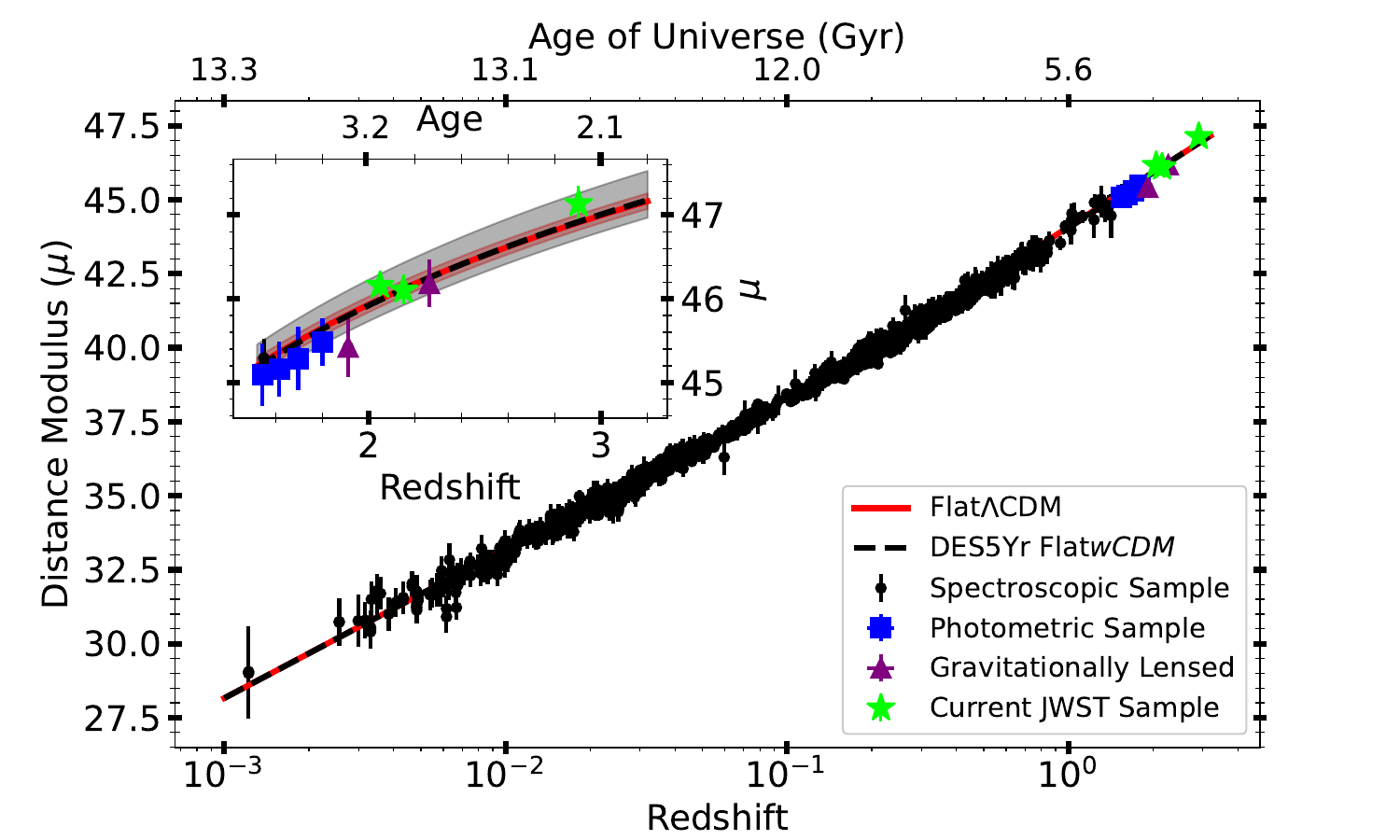}
    \caption{Luminosity distance measurements from the full sample of SNe Ia from \citet{brout_pantheon_2022} extending to $z = 2.22$. Black points (with errors) are SNe\,Ia
with spectroscopic classifications, while blue squares (with errors) are SNe Ia with photometric classifications. The two strongly lensed SNe Ia with distance measurements are shown as purple triangles. Green stars are the current high-$z$ \textit{JWST} SN\,Ia sample \citep{pierel_discovery_2024,pierel_testing_2025,siebert_sn_2025}. $\Lambda$CDM is shown as a solid red line (with shading representing $H_0$ values spanning the tension), and the best-fit Flat$w$CDM cosmological constraints from \citet[][black dashed line]{des_collaboration_dark_2024} is also shown for comparison.}
    \label{fig:ia_cosmo}
\end{figure}

SNe\,Ia have now been used for decades as precise luminosity distance measures to $z\sim2$, enabling the discovery of dark energy and our best local measurement of the Hubble constant \citep{riess_observational_1998,perlmutter_measurements_1999,riess_comprehensive_2022}. The exact nature of dark energy is still one of the most confounding mysteries in astrophysics today, and the upcoming \textit{Nancy Grace Roman Space Telescope} SN\,Ia cosmology program will seek to shed light on its properties. One open question that could impact dark energy measurements is if SN\,Ia standardization at high-$z$ is consistent with low-$z$, as anomalous distance measurements could indicate evolving dark energy and/or redshift-dependent SN\,Ia luminosity biases. Disentangling these effects is extremely difficult in the dark-energy-dominated universe ($z\lesssim1.5$), but beyond $z\sim2$, dark energy is vanishingly small and the Universe's expansion history is simple.  SNe\,Ia should closely follow the expected distance-redshift relation in this range, and any deviations could not be caused by dark energy and thus would require differences in the SNe themselves \citep{riess_first_2006}. \textit{JWST} has provided several of the first spectroscopically confirmed SNe\,Ia at $z\gtrsim2$, enabling an initial glimpse into the presence of any redshift-dependent luminosity biases \citep[][and see Figure \ref{fig:ia_cosmo}]{pierel_discovery_2024,pierel_testing_2025,siebert_sn_2025}. Thus far, while a small deviation from $\Lambda$CDM has been seen at $z=2.9$ \citep{pierel_discovery_2024}, the high-$z$ SNe\,Ia are consistent with current cosmological models that include evolving dark energy \citep{des_collaboration_dark_2024}. Therefore the existing sample suggests no extreme evolution in SNe\,Ia that could be significantly biasing existing cosmological measurements, though a larger sample at $z>2$ will be needed to limit such systematics to below the level required for the \textit{Nancy Grace Roman Space Telescope}. 

\section{New Insights into the Core-Collapse Supernova Population}\label{sec:cc}

\subsection{Building samples for population inferences}
\label{sub:cc-samples}
\begin{figure}[h!]
    \centering
    \includegraphics[width=\linewidth]{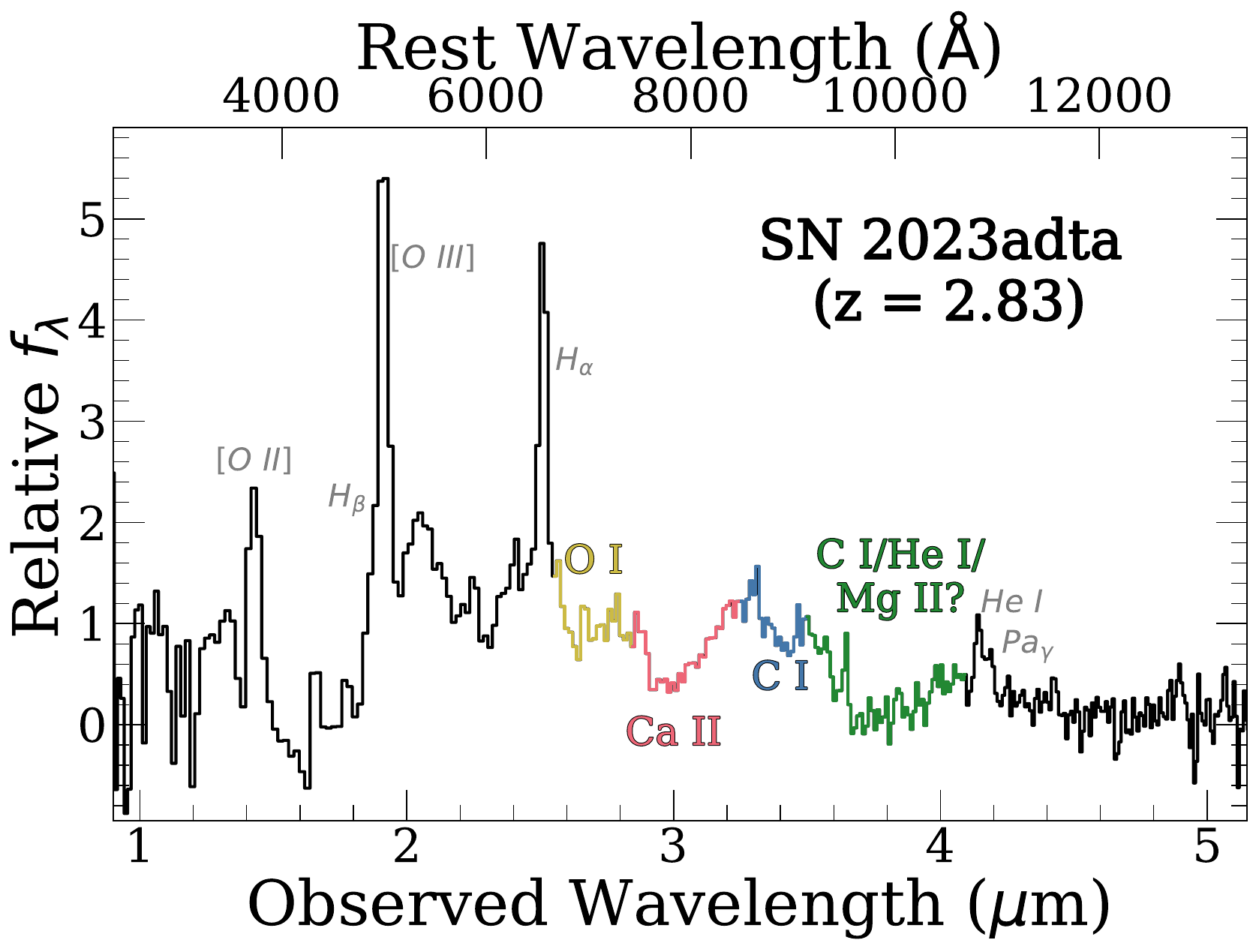}
    \caption{The Type Ic-BL SN (SN\,2023adta) discovered at $z$\,$=$\,2.83 in the JADES Transient Survey \citep{siebert_discovery_2024}. SNe\,Ic-BL represent just 4\% of the local  rate of core-collapse SNe by volume \citep{shivvers2017_sesn_fractions}, so it is interesting that an SN\,Ic-BL was immediately discovered at high-$z$ in a relatively small survey area. Figure adapted from \citet{siebert_discovery_2024}.}
    \label{fig:jades_ic-bl}
\end{figure}

\footnotetext[1]{\href{https://www.stsci.edu/jwst/science-execution/program-information.html?targetUrl=\%2Fjwst-program-info\%2Fprogram\%2F\%3Fprogram\%3D9478\%26pi\%3D1\%26referrer\%3Dhttps\%3A\%2F\%2Fwww.stsci.edu}{SN\,Ares \textit{JWST} program}}
\footnotetext[2]{\href{https://www.stsci.edu/jwst/science-execution/program-information.html?targetUrl=\%2Fjwst-program-info\%2Fprogram\%2F\%3Fprogram\%3D4568\%26pi\%3D1\%26referrer\%3Dhttps\%3A\%2F\%2Fwww.stsci.edu}{SN\,Athena \textit{JWST} program}}
The environmental conditions of the distant ($z$\,$\gtrsim$\,2) Universe differ drastically from the conditions found in the local Universe (e.g., lower metallicity). It is therefore plausible that the general properties of SN progenitors (and SNe themselves) at high-$z$ are significantly different from those at low-$z$. It is important to build large samples of distant SNe to answer questions such as: How do SN and host properties differ at high-$z$ and low-$z$? How do CC\,SN rates evolve with redshift? Are certain CC\,SN subtypes more common at high-$z$ than low-$z$? In the first three \textit{JWST} cycles, high-redshift SN science was largely conducted through high-redshift galaxy programs \citep[e.g.,][]{dunlop_primer_2021,casey_cosmos-web_2023,eisenstein_jades_2023,yan_pearls_2025}. One such program was the \textit{JWST} Advanced Deep Extragalactic Survey (JADES) program (\#1180; PI: Eisenstein; \citep{eisenstein_jades_2023}). JADES obtained two sets of deep ($\sim$30~mag) 9-band NIRCam images covering a $\sim$25 arcmin$^2$ portion of GOODS-S (the ``JADES Deep Field"), separated by one observer-frame year. The JADES Transient Survey was conducted by differencing these two sets of images, leading to the discovery of 79 SNe out to $z$\,$=$\,4.82 \citep{decoursey_jades_2024}. Remarkably, nearly
half of the sample (38 SNe) is at $z$\,$>$\,2, with 23 at $z$\,$=$\,2--3, 8 at $z$\,$=$\,3--4, and 7 at $z$\,$=$\,4--5. A subset of these SNe were spectroscopically classified via director's discretionary (DDT) program \#6541 (PI: Egami; \citep{egami_jwst_2023}), including a broad-line Type Ic SN at $z$\,$=$\,2.83 (Fig \ref{fig:jades_ic-bl}; \citep{siebert_discovery_2024}) and an SN IIP at $z$\,$=$\,3.61 \citep{coulter_discovery_2025}. The rapid discovery of a high-$z$ SN Ic broadline (Ic-BL) in a relatively small survey area may imply that SNe\,Ic-BL are more common at high-$z$, and the high explosion energies measured for a subset of the SNe\,II in the JADES SN sample provides an early indication that SN\,II explosion energies may increase with redshift \citep{moriya_properties_2025}. However, larger CC\,SN samples are clearly necessary to verify these early findings. The Public Release IMaging for Extragalactic Research (PRIMER; \#1837; PI: Dunlop; \citep{dunlop_primer_2021}), COSMOS-Web (\#1727; PI: Casey; \citep{casey_cosmos-web_2023}) and COSMOS-3D (\#5893; PI: Kakiichi) are additional extragalactic \textit{JWST} programs that were used to discover high-redshift SNe in the COSMOS field. Hundreds of SNe were discovered in the 0.037 deg$^2$ overlap between COSMOS-Web and PRIMER and the 0.54 deg$^2$ overlap between COSMOS-Web and COSMOS-3D (e.g., \citep{coulter2024_cweb_tns, coulter2025_c3d_tns, tee2025_c3d_tns}), including a bright, red, unusual CC\,SN at $z$\,$=$\,3 \citep{padilla-gonzalez2025_z3_proposal} and a likely dust-producing SN\,IIn at $z$\,$=$\,0.996 \citep{wang2025_blt_proposal}. 

There is an ongoing Cycle 4--6 \textit{JWST} program revisiting the JADES Deep Field with the primary goal of discovering and characterizing high-$z$ SNe (\#8060; PI: Egami; \citep{egami2025_jets}). Expected to discover $\sim$90 SNe ($\sim$45 at $z$\,$>$\,2) and spectroscopically classify $\sim$30 SNe, this program will provide the first definitive measurement of CC\,SN rates at $z$\,$>$\,2 and significantly expand the sample of high-$z$ CC\,SN spectra. With a large sample of spectroscopically-classified high-$z$ SNe, we can measure the general properties of high-$z$ SNe (e.g., explosion energies, rates) and compare those to their low-$z$ counterparts for the first time.

\subsection{Zooming in on a Core-Collapse Supernova at $z=5.13$}
\label{sub:highz-eos}
\begin{figure}[h!]
    \centering
    \includegraphics[width=\linewidth]{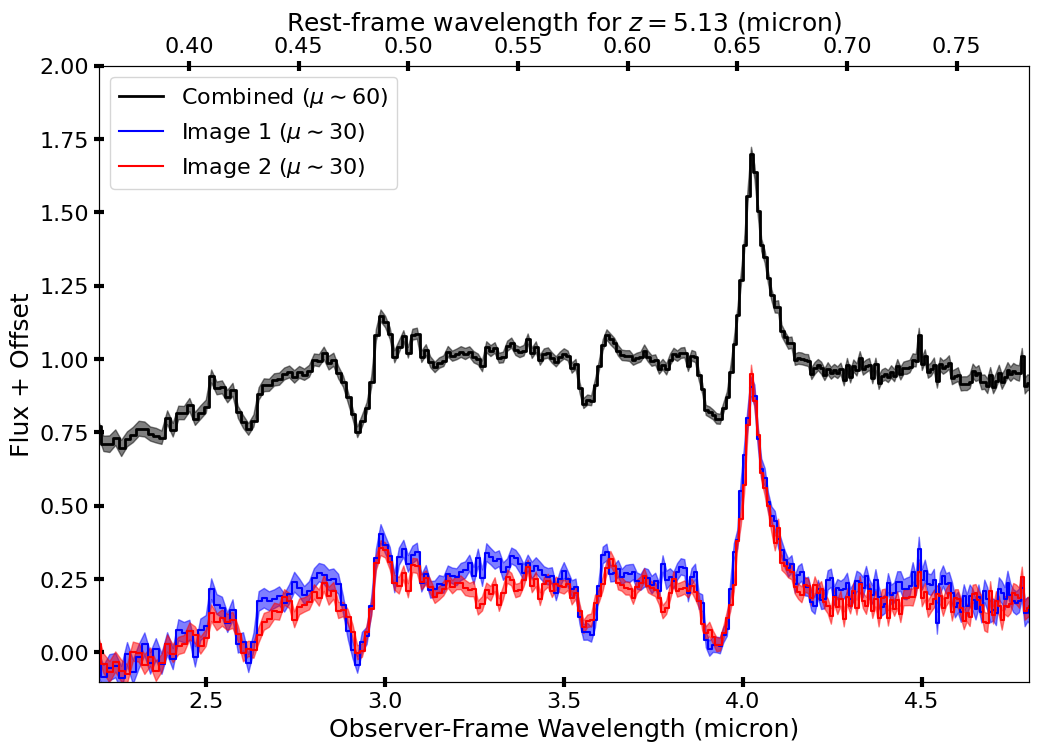}
    \caption{The SN~Eos\,spectra for the two visible images of SN~Eos (red and blue) are shown with both observer- and rest-frame axes. A co-added spectrum is shown in black, as the phases of the two images are expected to be within $\sim3$ rest-frame days of each other. The spectrum is remarkably well-matched by normal SN\,IIP spectra in the local universe, with the exception of Fe\,II lines related to metallicity \citep{coulter_spectroscopically_2026}. Figure adapted from \citet{coulter_spectroscopically_2026}.}
    \label{fig:eos_spec}
\end{figure}

While \textit{JWST} has significantly expanded our view of CC\,SNe at $z>1$ relative to \textit{HST} \citep[e.g.,][]{strolger_rate_2015}, spectroscopic classification has been limited to $z\sim3.6$ for non-lensed SNe \citep{coulter_discovery_2025}. High S/N ($\gtrsim10$) spectroscopy has only been possible at $z<3$ thus far, and only for a handful of objects \citep{siebert_discovery_2024,siebert_sn_2025}. The large magnifications afforded by galaxy cluster lensing systems can greatly increase the signal of very distant sources, providing rare opportunities to study extremely distant SNe in extraordinary detail. SN~Eos, a doubly-imaged, strongly lensed transient is the highest-redshift ($z=5.13$) SN yet spectroscopically classified. Analysis of the spectrum and light curve of SN~Eos revealed that it is of the type IIP sub-class (Figure \ref{fig:eos_spec}), a SN originating from the collapse of a massive star with $Z\lesssim0.1Z_{\odot}$ \citep{coulter_spectroscopically_2026}. Continued observations of galaxy clusters \citep[e.g.,][]{fujimoto_vast_2025} may yet reveal additional extremely high-$z$ SNe in the same remarkable detail as SN~Eos, providing rare insights into stellar populations in the early universe.

\section{Summary and Discussion}\label{sec:discussion}
We have presented an overview of SN discoveries and scientific breakthroughs of the first few years of \textit{JWST} observations. Nearby SN efforts have primarily been leveraging the first opportunity to observe SNe with deep, high-resolution, mid-IR imaging and spectroscopy far beyond the capabilities of \textit{Spitzer}. Such efforts have led to insights into dust production in CC\,SNe, which could account for large dust reservoirs in the early universe, as well as some of the first mid-IR nebular phase spectroscopy for SNe\,Ia. These are an extremely promising path to understanding SN\,Ia progenitor systems, a long-standing question with implications for SN\,Ia cosmology. At higher redshifts, \textit{JWST} has already made contributions to cosmology at large with two new measurements of $H_0$ using strongly lensed SNe, and constraints on redshift-dependent SN\,Ia luminosity biases that (if present) could strongly impact future dark energy measurements. Both of these topics have a bright future with \textit{JWST}. \textit{JWST}-observed lensed SN $H_0$ measurements should be able to make a $\sim3\sigma$ statement about the $H_0$ tension in the near future, and the sample of $z>2$ SNe\,Ia should be able to constrain luminosity biases to below the level of intrinsic SN\,Ia dispersion in advance of \textit{Roman}. Finally, \textit{JWST} has provided new constraints on CC\,SNe through surveys building large samples of SNe for population level inferences and novel rate measurements, as well as detailed views (with the aid of gravitational lensing) of the most distant CC\,SN yet observed. 

The \textit{Roman Space Telescope}, expected to launch in mid-$2026$, will execute the high latitude time-domain survey (HLTDS) as one of its core community surveys. The HLTDS will be the widest area, deepest near-IR time-domain survey yet, and is expected to be transformative for a wide range of science cases. In particular, the HLTDS will be an excellent source of cosmologically useful lensed SNe \citep{pierel_projected_2021}, high-$z$ SNe \citep{hounsell_simulations_2018,rose_host_2021}, and rest-frame near-IR observations of low-$z$ SNe \citep[e.g.,][]{pierel_salt3nir_2022}. However, spectroscopy for this sample will be limited, with \textit{Roman} itself only providing slitless spectroscopy, which will not be sufficient for SNe that are deeply embedded in their host galaxy, intrinsically faint, or very distant. Due to a combination of sensitivity and wavelength limitations, \textit{Roman} therefore will still rely on \textit{JWST} to robustly characterize objects in these scenarios with observer-frame IR/mid-IR imaging and spectroscopy. Therefore while \textit{Roman} represents a revolution for a broad range of time-domain science, \textit{JWST} will continue to play a key complementary role into the future as it continues to build unique datasets for a diverse sample of SNe. 

\bmhead{Acknowledgements}

We thank Claudio Grillo and Sherry Suyu for organizing and editing this special issue. 

\bmhead{Declarations}
\begin{enumerate}
    \item \textbf{Funding:} J.P. is supported by the National Aeronautics and Space Administration (NASA) through Einstein Fellowship grant  HF2-51541.001 awarded by the Space Telescope Science Institute (STScI), which is operated for NASA by the Association of Universities for Research in Astronomy, Inc.,  under contract NAS5-26555. The high-z SNe research was supported by \textit{JWST} program grants JWST-GO-05324, JWST-GO-08060, JWST-GO-09493, JWST-GO-12299, JWST-GO-12146, and NASA EGIP 24-EGIP24-0034. D.A.C. acknowledges funding through \textit{JWST} program grants JWST-GO-06541, JWST-GO-06585, and JWST-GO-05324. The work related to SN dust was supported by JWST-GO-1860, JWST-GO-2666, JWST-GO-3921, JWST-GO-6356, JWST-GO-8878, JWST-GO-8883, JWST-GO-12264. M.S. is supported by NASA through an STScI Postdoctoral Fellowship awarded by STScI. J.D. is supported in part by NASA grants JWST-GO-02114, JWST-GO-03726, JWST-GO-6023, JWST-GO-6582, and HST-AR-17555. MG and ME were supported in part by NASA grants JWST-GO-5324. CL was supported through NASA grants JWST-GO-4568, HST-GO-17474, and JWST-GO-9478. E.P.G. is supported by NSF AAG grant 2205635. Support for program  Nos. 2114, 3726, 6023, 6582, 6677, and 17555 were provided by NASA  through grants from the Space Telescope Science Institute, which is  operated by the Association of Universities for Research in Astronomy, Inc., under NASA contract NAS 5-03127.

    \item \textbf{Data availability:} All data referred to in this paper are available to the public on the Mikulski Archive for Space Telescopes (MAST), \url{https://mast.stsci.edu/portal/Mashup/Clients/Mast/Portal.html}
    \item \textbf{Competing Interests:} The authors have no relevant financial or non-financial interests to disclose.

    \item \textbf{Authors Contributions:} All authors contributed to the concept of the paper, and to their relevant descriptions of science accomplished/planned with \textit{JWST}. The first draft of the manuscript was written by Justin Pierel, David Coulter, Christa DeCoursey, James DerKacy, Ori Fox, Melissa Shahbandeh, and Matthew Siebert, and all authors commented on previous versions of the manuscript. All authors read and approved the final manuscript.
\end{enumerate}

\clearpage

\bibliographystyle{sn-nature}

\end{document}